\documentclass[aps,prb,twocolumn,showpacs,superscriptaddress,longbibliography]{revtex4-2}  

\usepackage{amssymb}
\usepackage{amsfonts}
\usepackage{amsmath}
\usepackage{bm}
\usepackage{textcomp}
\usepackage{color}
\usepackage{graphicx}
\usepackage{float}

\usepackage[a4paper=true,pagebackref=false]{hyperref}

\usepackage{tikz,xcolor,hyperref}

\definecolor{lime}{HTML}{A6CE39}
\DeclareRobustCommand{\orcidicon}{%
	\begin{tikzpicture}
	\draw[lime, fill=lime] (0,0)
	circle [radius=0.16]
	node[white] {{\fontfamily{qag}\selectfont \tiny ID}};
	\draw[white, fill=white] (-0.0625,0.095)
	circle [radius=0.007];
	\end{tikzpicture}
	\hspace{-2mm}
}

\foreach \x in {A, ..., Z}{%
	\expandafter\xdef\csname orcid\x\endcsname{\noexpand\href{https://orcid.org/\csname orcidauthor\x\endcsname}{\noexpand\orcidicon}}
}

\begin{document}

\title[PEME]{Electric-field manipulation of magnetization in an insulating dilute ferromagnet through piezoelectromagnetic coupling}

\author{Dariusz~Sztenkiel\orcidA} \email{sztenkiel@ifpan.edu.pl}
\affiliation{Institute of Physics, Polish Academy of Sciences, Aleja Lotnikow 32/46, PL-02668 Warsaw, Poland}

\author{Katarzyna~Gas\orcidD}
\affiliation{Institute of Physics, Polish Academy of Sciences, Aleja Lotnikow 32/46, PL-02668 Warsaw, Poland}
\affiliation{Center for Science and Innovation in Spintronics, Tohoku University, Katahira 2-1-1, Aoba-ku, Sendai 980-8577, Japan}

\author{Nevill~Gonzalez Szwacki\orcidG}
\affiliation{Faculty of Physics, University of Warsaw, ul.\,Pasteura 5, PL-02093 Warszawa, Poland}

\author{Marek~Foltyn\orcidF}
\affiliation{Institute of Physics, Polish Academy of Sciences, Aleja Lotnikow 32/46, PL-02668 Warsaw, Poland}
\affiliation{International Research Centre MagTop, Institute of Physics, Polish Academy of Sciences,
Aleja Lotnikow 32/46, PL-02668 Warsaw, Poland}

\author{Cezary~\'Sliwa\orcidJ}
\affiliation{International Research Centre MagTop, Institute of Physics, Polish Academy of Sciences,
Aleja Lotnikow 32/46, PL-02668 Warsaw, Poland}

\author{Tomasz~Wojciechowski\orcidI}
\affiliation{International Research Centre MagTop, Institute of Physics, Polish Academy of Sciences, Aleja Lotnikow 32/46, PL-02668 Warsaw, Poland}

\author{Jaros\l{}aw~Z.~Domagala\orcidH}
\affiliation{Institute of Physics, Polish Academy of Sciences, Aleja Lotnikow 32/46, PL-02668 Warsaw, Poland}

\author{Detlef~Hommel}
\affiliation{Polish Center of Technology Development, ul.\,Stab{\l}owicka 147, PL 54-066, Wroc{\l}aw, Poland}

\author{Maciej~Sawicki\orcidB} \email{sawicki@ifpan.edu.pl}
\affiliation{Institute of Physics, Polish Academy of Sciences, Aleja Lotnikow 32/46, PL-02668 Warsaw, Poland}
\affiliation{Research Institute of Electrical Communication, Tohoku University, Katahira 2-1-1, Aoba-ku, Sendai 980-8577, Japan}

\author{Tomasz~Dietl\orcidC} \email{dietl@MagTop.ifpan.edu.pl}
\affiliation{International Research Centre MagTop, Institute of Physics, Polish Academy of Sciences,
Aleja Lotnikow 32/46, PL-02668 Warsaw, Poland}


\begin{abstract}

The electric field control of magnetization is of significant interest in materials science due to  potential applications in many devices such as sensors, actuators,  and magnetic memories. 
Here, we report magnetization changes generated by an electric field in ferromagnetic Ga$_{1-x}$Mn$_x$N grown
by molecular beam epitaxy. Two classes of phenomena have been revealed. First, over a wide
range of magnetic fields, the magnetoelectric signal is odd in the electric field and
reversible. Employing a macroscopic spin model and atomistic Landau-Lifshitz-Gilbert theory with
Langevin dynamics, we demonstrate that the magnetoelectric response results from
the inverse piezoelectric effect that changes the trigonal single-ion magnetocrystalline anisotropy.
Second, in the metastable regime of ferromagnetic
hystereses, the magnetoelectric effect becomes non-linear and irreversible in response to
a time-dependent electric field, which can reorient the magnetization direction. Interestingly, our
observations are similar to those reported for another dilute ferromagnetic semiconductor
Cr$_x$(Bi$_{1-y}$Sb$_y$)$_{1-x}$Te$_3$, in which magnetization was monitored as a function of the gate electric
field. Those results constitute experimental support for theories describing the effects of
time-dependent perturbation upon glasses far from thermal equilibrium in terms of an enhanced
effective temperature.

\end{abstract}

\maketitle

\baselineskip 15pt

\section*{Introduction}


Repetitive switching of magnetization in ferromagnetic materials is a key process in magnetic recording and information storage. Traditionally, this energy-intensive process is achieved by applying an external magnetic field or spin-polarized currents. An emerging alternative, however, is the use of an electric field to control magnetic anisotropy, which could offer a more energy-efficient solution compared to conventional methods. Different approaches have been proposed in the past to observe the magnetoelectric (ME) effect, in which coupling between magnetism and electricity is present, such as the gate-controlled hole concentration and the Curie temperature in dilute ferromagnetic semiconductors\cite{Ohno:2000_N,Boukari:2002_PRL,Chiba:2003_S,Dietl:2014_RMP}, voltage induced structural and magnetic phase transition\cite{Cui:2017_PRAppl, Yi:2020_NatComm}, piezoelectric based modification of magnetism\cite{Casiraghi:2012_APL,Sztenkiel:2016_NatComm,Arora:2019_PRM} or magnetoelectric effect in multiferroics\cite{Molinari:2019_AdvMat,Ramesh:2008}. The ME effect has  been observed  in ferromagnetic\cite{Casiraghi:2012_APL}, antiferromagnetic\cite{Liu:2016_PRL,Lee:2015_NatComm},  superparamagnetic\cite{Arora:2019_PRM}, and paramagnetic\cite{Sztenkiel:2016_NatComm} materials.

A first step to accomplish the repetitive switching of magnetization is the demonstration of the control of the key magnetic properties, such as the coercive field, saturation magnetization, or remanent magnetization via electric fields\cite{Chiba:2003_S,Pertsev:2010_Nanotechnology,Brandlmaier:2011_JAP}.	A strong coupling between an electric field and magnetism was shown to exist and theoretically quantified in piezoelectric wurtzite (wz) Ga$_{1-x}$Mn$_x$N in the paramagnetic phase, that is for Mn concentrations $x \lesssim 2.5$\% and temperatures $T \geq 2$~K (ref.~\onlinecite{Sztenkiel:2016_NatComm}). Those samples were grown by metalorganic vapour-phase epitaxy (MOVPE)\cite{Bonanni:2011_PRB}. Now aided by a carefully elaborated molecular beam epitaxy (MBE) technique,  we incorporate substitutionally up to 10$\%$  of Mn ions into GaN host\cite{Kunert:2012_APL,Gas:2018_JALCOM}. In those semi-insulating layers, Mn$^{3+}$ ions are coupled  by a short-range ferromagnetic (FM) superexchange\cite{Bonanni:2011_PRB,Sawicki:2012_PRB,Stefanowicz:2013_PRB,Sliwa:2024_PRB}, and  FM ordering is realized in a percolation-like fashion\cite{Korenblit:1973_PLA,Bergqvist:2004_PRL,Bonanni:2021_HB}.

Our combine experimental and theoretical studies of the ME effect in the ferromagnetic (Ga,Mn)N address two classes of questions. First, we demonstrate that the sign and magnitude of magnetization generated by an electric field, $M_E$, we find by superconducting quantum interference device (SQUID) magnetometry, result from a change of uniaxial magnetic anisotropy driven by the inverse piezoelectric effect. Our conclusion is supported by  two  complementary theoretical approaches: (i) a macrospin model, in which changes of magnetic anisotropy are described in terms of magnetizing or demagnetizing anisotropy fields, depending on the direction of the external magnetic field with respect to the easy axis; (ii) an atomistic Landau-Lifshitz-Gilbert approach for interacting classical spins\cite{Skubic:2008_JPhysCM,Evans:2014_JPhysCM,Evans:2015_PRB} with input parameters obtained by fitting our magnetization $M(H)$ data. Our results, on the one hand imply that the ME effect can be understood and designed in well characterized magnetic systems. On the other, we find that a short time scale inherent to the LLG approach leads to an overestimation in the theoretical magnitudes  of the remanent  magnetization and magnetoelectric effect in dilute ferromagnets in the magnetic fields $H \rightarrow 0$.

Second, the percolating ferromagnetism means the presence of an infinite FM cluster at $T \le T_{\text{C}}$. This cluster encompasses all spins at $T =0$ but only about 20\% of spins  at $T$ approaching $T_{\text{C}}$ (ref.~\onlinecite{Korenblit:1973_PLA}).  Thus, at any non-zero temperature, a large portion of the Mn ions resides in {\em finite} clusters of different sizes. Such a magnetic system is expected to exhibit a broad spectrum of relaxation times and, in the hysteretic regime,  constitutes a glass far from thermal equilibrium. Thus, a possibility of perturbing ferromagnetic (Ga,Mn)N by an electric field offers a unique opportunity to test theoretical predictions concerning dynamics of driven non-equilibrium glass systems\cite{Makse:2002_N,Loi:2008_PRE,Berthier:2013_NP}.

As we show, our data provide a strong support for persisting theoretical proposals arguing that a non-thermal and non-dissipative drive, bringing a glass beyond the linear response regime, can be described by enlarged effective temperature\cite{Berthier:2013_NP}. Interestingly, this interpretation appears explaining surprising magnetization behaviour observed in  dilute ferromagnetic semiconductor (Bi,Sb,Cr)$_2$Te$_3$ in the regime of the quantum anomalous Hall (QAH) effect\cite{Lachman:2015_SA}. Tapping of that QHE ferromagnet by gate voltage modulation resulted, according to local SQUID measurements, in magnetization shift towards the direction of the external magnetic field in the hysteretic regime\cite{Lachman:2015_SA}, as reported here for (Ga,Mn)N. As short-range ferromagnetic superexchange dominates also in topological ferromagnetic insulators\cite{Sliwa:2021_PRB}, we expect similar glassy-like relaxation towards thermal equilibrium under a time-dependent drive in both systems.

It is worth noting, however, that our conclusions should not be extended towards experiments, in which magnetization switching in dilute ferromagnetic semiconductors was revealed by resistance measurements\cite{Wurstbauer:2010_NP,Liu:2016_SA}, as an important additional ingredient there is Joule heating that may lead to surprising bistable behaviours\cite{Wurstbauer:2010_NP}. Still another physics may show up in {\em mesoscopic} QAH samples, in which resistance switching was found also  in  magnetic fields far above the coercivity field\cite{Grauer:2015_PRB,Fijalkowski:2023_SA}. In such submicron wires, a highly probable process of backscattering between edge states is controlled by (i) orientation of local magnetization and (ii) spatial arrangement of localized in-gap carriers that are frozen far from thermal equilibrium under the presence of the Hall electric field. Resistance switching dynamics was found to accelerate with the magnetic field in the ferromagnetic QAH case\cite{Fijalkowski:2023_SA}, as found earlier in mesoscopic {\em antiferromagnetic} spin-glasses\cite{Jaroszynski:1998_PRL}. It is to be seen on whether  the AC source-drain voltage in QAH nanostructures\cite{Fijalkowski:2023_SA} just allowed tracing magnetization dynamics or rather had an effect on the dynamics and effective temperature of the {\em electron} glass.

In our work we demonstrate a clear control of magnetic anisotropy of piezoelectric and ferromagnetic (Ga,Mn)N layers by application of an electric field. We observe a reduction of the width of a hysteresis curve from 120~Oe to about 10~Oe as the magnitude of the applied AC voltage $V_{\text{AC}}$ increases from 0.2~V to 2~V and a non-reversible switching of magnetization toward the equilibrium  direction for magnetic fields close to the coercive field, occurring under a stepped change of the electric field either elongating or shortening the lattice parameter $c$.

\section*{Results and Discussion}

\ \\
{\bf Samples}. Two samples with different thicknesses of (Ga,Mn)N layer $d_{\mathrm{(Ga,Mn)N}}$ and similar concentrations of Mn ions $x$ are investigated here, namely, $d_{\mathrm{(Ga,Mn)N}}=10$~nm, $x=5.6\%$ (Sample A) and $d_{\mathrm{(Ga,Mn)N}}=160$~nm, $x=6\%$ (Sample B). They were grown along the $c$ axis on a thick GaN buffer followed by an $n^+$-GaN:Si layer ($n \sim 3\times10^{18}$\,cm$^{-3}$), which serves as a back-contact contact layer, and characterized according to protocols elaborated previously\cite{Kunert:2012_APL,Gas:2018_JALCOM}.  The experimental findings\cite{Sawicki:2012_PRB,Stefanowicz:2013_PRB} and the theoretical modeling\cite{Simserides:2014_EPJ} agrees that for $ x < 10$\% Curie temperature $T_{\mathrm{C}}$ of Ga$_{1-x}$Mn$_x$N obeys the power law $T_{\mathrm{C}}$~$\propto x^{2.2}$, which well approximates the percolation exponential formula for small $x$ values. The layers studied here do fall into this dependency exhibiting $7\lesssim T_{\mathrm{C}} \lesssim 8$~K for $x \simeq 6$\%.

\begin{figure}[tbh]
\centering
\includegraphics[width=8.5 cm]{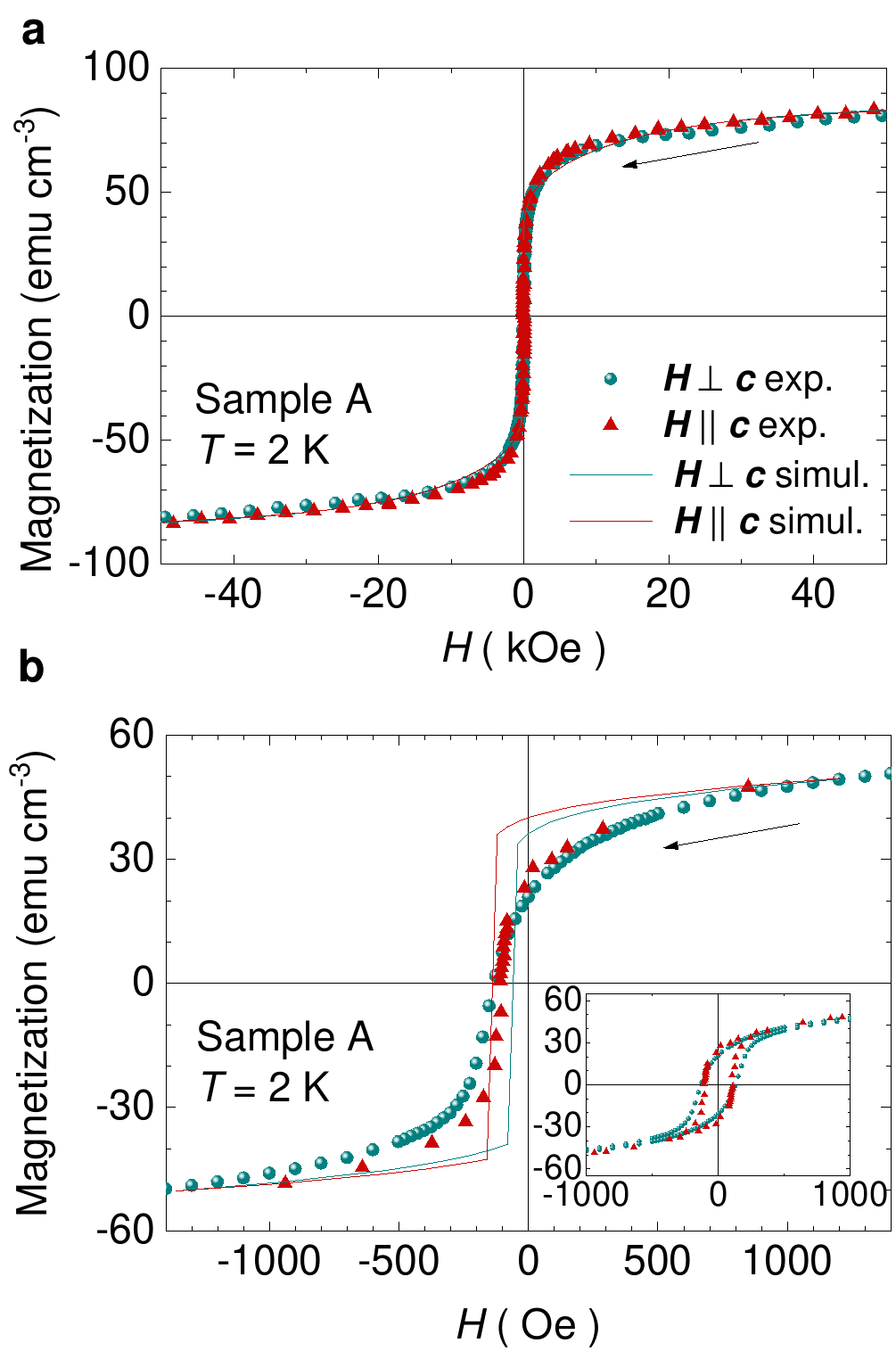}
\caption{\label{Fig:Magnetization} {\bf Magnetization, Sample A.} {\bf a} and {\bf b}  Full points: experimental magnetization $M$ of Ga$_{1-x}$Mn$_{x}$N with $x=5.6\%$ measured at $T = 2$~K for two perpendicular magnetic field orientations, in-plane ($\textbf{H} \bot \textbf{c}$) and out-of-plane ($\textbf{H} \| \textbf{c}$). Shape anisotropy (demagnetization) effects are removed from experimental data. Inset in {\bf b}, Close-up on the weak magnetic field regime ($H$ in Oe). Lines in {\bf a} and {\bf b}: magnetization calculated for magnetic field sweep direction shown by arrow using atomistic Landau-Lifshitz-Gilbert equations with Langevin dynamics for 9464 classical spins with single-ion magnetocrystalline anisotropy and Heisenberg ferromagnetic interactions between spins (up to 18$^{th}$ neighbours), as described in the theoretical subsection.}
\end{figure}

In Fig.~\ref{Fig:Magnetization} we compare $M(H)$ data of Sample A acquired at $T=2$~K for both in-plane ($\textbf{H} \bot \textbf{c}$) and out-of-plane ($\textbf{H} \| \textbf{c}$) configuration for which the demagnetization field was subtracted from the external field value. We see that at that temperature $T \approx T_{\text{C}}/4$ and the time scale of the experiment, magnetic hysteresis involves only a part of spins, substantiating expectations of the percolation model\cite{Korenblit:1973_PLA,Bergqvist:2004_PRL,Bonanni:2021_HB}.

The data indicate also that magnetocrystalline anisotropy is close to zero for the Mn content and strain in question. In this case, shape anisotropy makes that the easy axis is in-plane in ferromagnetic thin films, as found previously\cite{Stefanowicz:2013_PRB,Gas:2018_JALCOM} and here. However, in samples with smaller $x$ values, uniaxial magnetocrystalline anisotropy is clearly non-zero and its sign corresponds to the in-plane easy axis\cite{Stefanowicz:2010_PRB,Sztenkiel:2016_NatComm}. Those data imply a variation with $x$ of the parameter $\xi=c/a-\sqrt{8/3}$, where $c$ and $a$ are lattice parameters of the wz unit cell. The parameter $\xi$ governs the trigonal deformation and the corresponding uniaxial magnetic anisotropy in magnetically doped wz crystals like Mn-doped GaN\cite{Gosk:2005_PRB,Stefanowicz:2010_PRB,Sztenkiel:2016_NatComm} or Co-doped ZnO\cite{Sawicki:2013_PRB}.


\begin{figure}[tbh]
\centering
\includegraphics[width=8.5 cm]{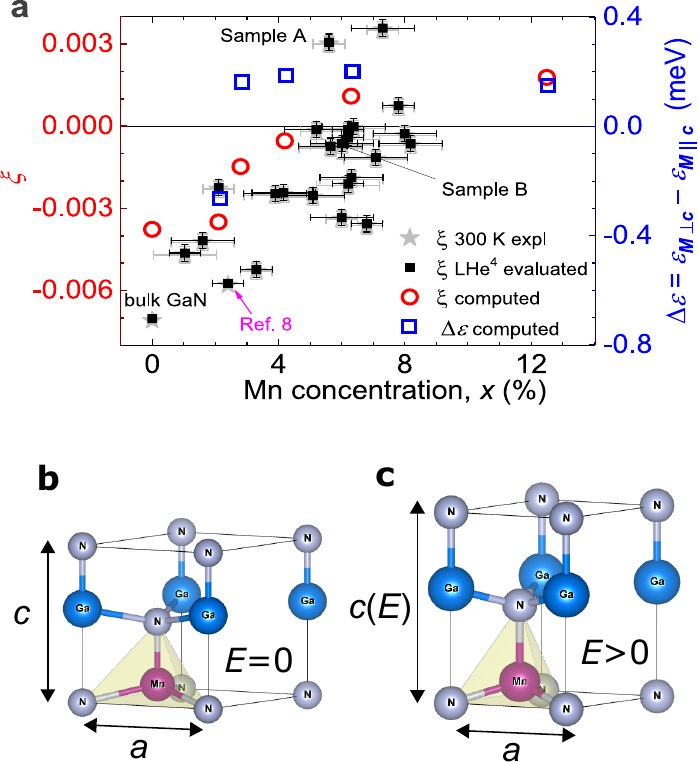}
\caption{\label{Fig:SamplePiezo} {\bf Wurtzite trigonal deformation.} {\bf{a}}  Left axis:  dependence of the parameter $\xi=c/a-\sqrt{8/3}$ on the Mn concentration $x$ measured experimentally at 300\,K (stars) essentially identical to evaluated values for 4.2\,K (full squares) compared to determined by {\em ab initio} computations for unstrained and optimized structure (open red circles). Right axis: total energy difference between the in plane (along the $a$-axis) and parallel to the $c$-axis configurations of the Mn magnetic moment in GaN (open blue squares). Labels indicate $\xi$ values for Samples A,  B and MOVPE film from ref.\,\onlinecite{Sztenkiel:2016_NatComm}.({\bf b}) and ({\bf c}) In Ga$_{1-x}$Mn$_{x}$N, the uniaxial magnetic anisotropy originates primarily from the trigonal deformation of the tetrahedron formed by anion atoms next to Mn. The application of an external electric field ${\textbf{E}}$ elongates lattice parameter $c$ by the inverse piezoelectric effect and controls uniaxial magnetocrystalline anisotropy.}
\end{figure}

To substantiate those findings, we show in Fig.~\ref{Fig:SamplePiezo}a $\xi$ values  obtained by $x$-ray diffraction for a series of Ga$_{1-x}$Mn$_{x}$N samples deposited by MOVPE and MBE on a GaN buffer clamping the $a$ value. We also present results of our first-principles calculations with the structure optimization (see Methods) by plotting the computed values of $\xi$ and the total energy difference $\Delta \varepsilon$ for (Ga,Mn)N with a Mn magnetic moment lying in a plane (along the \textit{a}-axis) and a Mn magnetic moment aligned with the \textit{c}-axis as a function of $x$. As seen in Fig.~\ref{Fig:SamplePiezo}a,  the energy difference changes the sign from negative to positive for the Mn concentration of about $2.5\%$, which correlates but is slightly lower than the computed value corresponding to the sign change of $\xi$. Those experimental and theoretical data imply that as a function of Mn content, and in the absence of the shape anisotropy, the Mn magnetic moment will have a tendency to reorient from the in-plane to perpendicular direction. Our two samples are close to that reorientation transition with $\xi =0.0029(3)$ (Sample A) and $-0.0005(3)$ (Sample B), in agreement with magnetization data in Fig.~\ref{Fig:Magnetization}.

\ \\
{\bf Experimental}. As presented in Methods, for magnetoelectric investigations Ga$_{1-x}$Mn$_x$N layer resides between the $n^+$-GaN:Si bottom contact and a top HfO$_2$ film covered by a metal gate. Such an arrangement  allows for  blocking gate currents resulting from a possible conductance $via$ threading dislocations in the (Ga,Mn)N epilayer. In the case of the DC voltage applied to the structure, the electric field on the HfO$_2$ and (Ga,Mn)N films is distributed according to relative resistances of the two layers, so that a rather small electric field is expected in (Ga,Mn)N samples as, presumably, $R_{\text{HfO2}} \gg  R_{\text{(Ga,Mn)N}}$. In contrast, for resistances and frequencies in question ($f = 17$\,Hz), the AC voltage is distributed according to relative capacitances, meaning that the application of the AC voltage of $V_{\text{AC}} = 1$\,V to the investigated structures (see Methods) corresponds
to the AC electric field $E_{\text{AC}}$ of about 0.19\,MV/cm and 0.04\,MV/cm to the (Ga,Mn)N layer of Sample
A and B, respectively. Given the structure parameters and wiring layout, we do not expect temperature changes by capacitor charging currents at experimental temperature of 2\,K.

The measurement system, depicted in Methods, utilizes a SQUID magnetometer and phase sensitive detection. An AC voltage $V_{\text{AC}}$ is applied between the metal gate and the GaN:Si contact layer. The resulting AC magnetization signal of the sample is picked up by SQUID coils and detected by a lock-in amplifier. Alternatively, we look for magnetization alterations generated by stepped DC voltage changes. As shown in  panels {\bf b} and {\bf c} of Fig.~\ref{Fig:SamplePiezo}, in our piezoelectric structure with the clamped value of $a$ by the thick buffer layer, an electric field $E$ alters only the lattice parameter $c$, according to $\Delta c/c=d_{33}E$, where $d_{33}  = 2.8$\,pm/V, as obtained for an epitaxial GaN film with the same orientation of the ${\textbf c}$ vector as in our samples, and also with a clamped value of $a$ by a thick GaN buffer layer\cite{Guy:1999_APL}.

\ \\
{\bf Experimental results and macrospin model}. As mentioned in the introduction, our data reveal the existence of two distinct magnetoelectric phenomena. First of them, discussed in this subsection, corresponds to the regime, where the magnetoelectric response $M_E$ is linear in the electric field $E$.  Given a rather unique combination of ferromagnetism and piezoelectricity offered by single crystals of wurtzite (Ga,Mn)N, the data in the linear range makes possible to test quantitatively our understanding of magnetoelectricity in dilute ferromagnets. The second phenomenon, presented in the last subsection of Results, is specified by strong nonlinearities and irreversibilities of $M_E$. As we show, the data in that regime give access to the physics of driven glassy systems far from thermal equilibrium.

We recall that in our experiments the electric field is applied along the $c$ axis of the wurtzite structure, and we measure magnetization changes in the direction of the applied magnetic field ${\textbf{H}}$.  We have found that in high magnetic fields, $M_E(H)$ is linear in $E$ up to the highest electric field supported by our devices. In contrast, in the hysteresis regime, nonlinearties are already apparent in the electric fields lower by a factor of ten. Thus, being interested here in the linear regime, we discuss $M_E$ values normalized to $V_{\text{AC}} =0.5$\,V (Sample A). Such a low value of $V_{\text{AC}}$ has been employed in low magnetic fields $H \leq 1$\,kOe but $V_{\text{AC}}$ has been risen up to 5\,V for $H \geq 10$\,kOe.

\begin{figure*}[htb]
\centering
\includegraphics[width=16 cm]{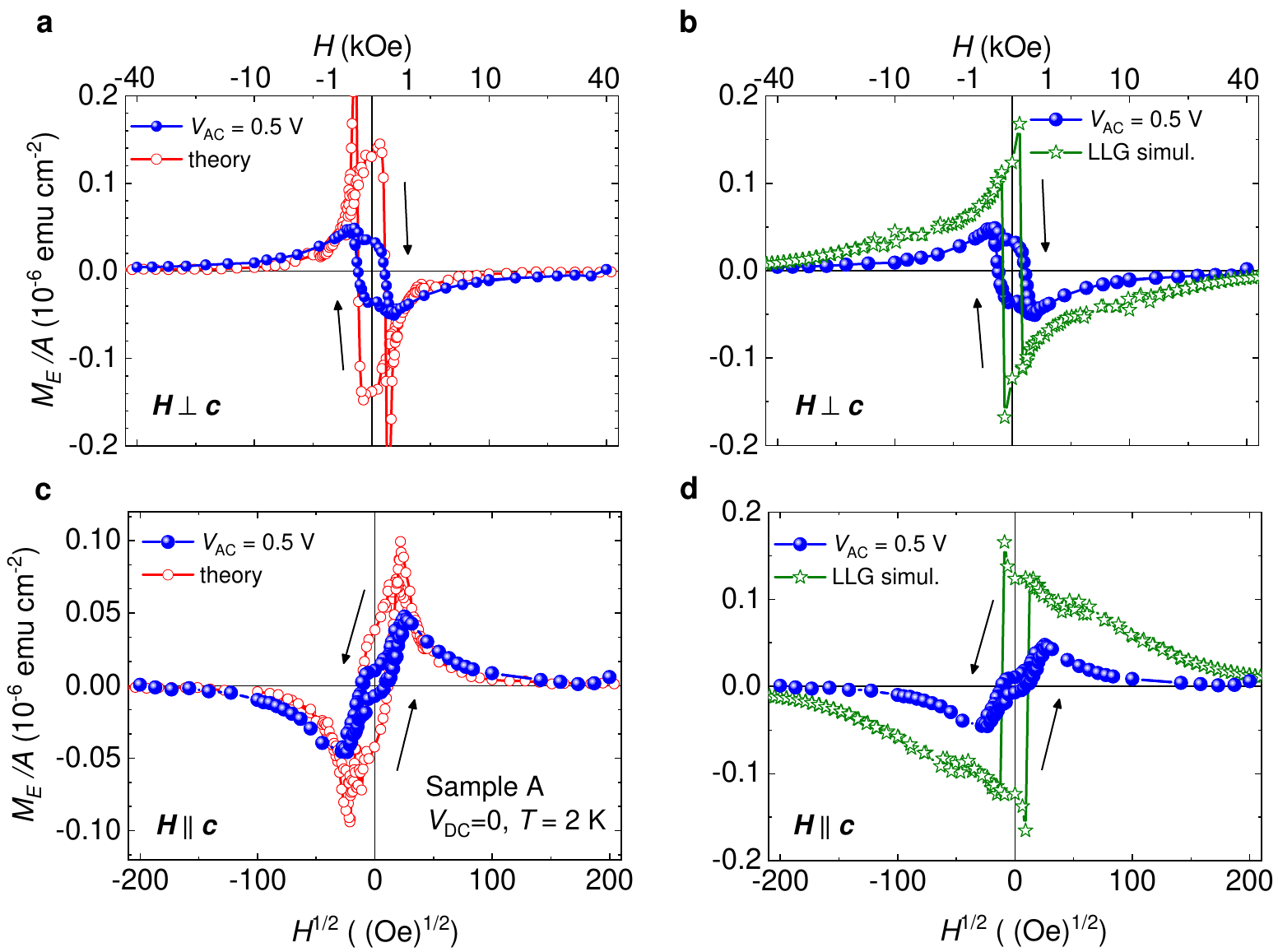}
\caption{\label{Fig:MEsimtheory} {\bf Magnetoelectric effect, Sample A.} Areal density of the magnetic moment $M_E/A$ normalized to the voltage $V_{\text{AC}}= 0.5$\,V in Ga$_{1-x}$Mn$_{x}$N layer with $x=5.6\%$  at $T = 2$\,K, i.e., in the ferromagnetic phase (blue solid points). The data are plotted as a function of the square root of the magnetic field to show simultaneously weak and strong field regions. Measurements are taken for the in-plane ($\textbf{H} \bot \textbf{c}$) and out-of-plane ($\textbf{H} || \textbf{c}$) configurations.  {\bf a} and {\bf c}  Open red points: theoretical results of magnetoelectric signal $M_E$ obtained from the macroscopic spin model (Eq.~\ref{eq:ME}) with one fitting parameter $b = 0.04$. {\bf b} and {\bf d} Open green stars: the magnetoelectric signal computed by using the atomistic Landau-Lifshitz-Gilbert (LLG) equations with Langevin dynamics. }
\end{figure*}

\begin{figure*}[htb]
\centering
\includegraphics[width=17 cm]{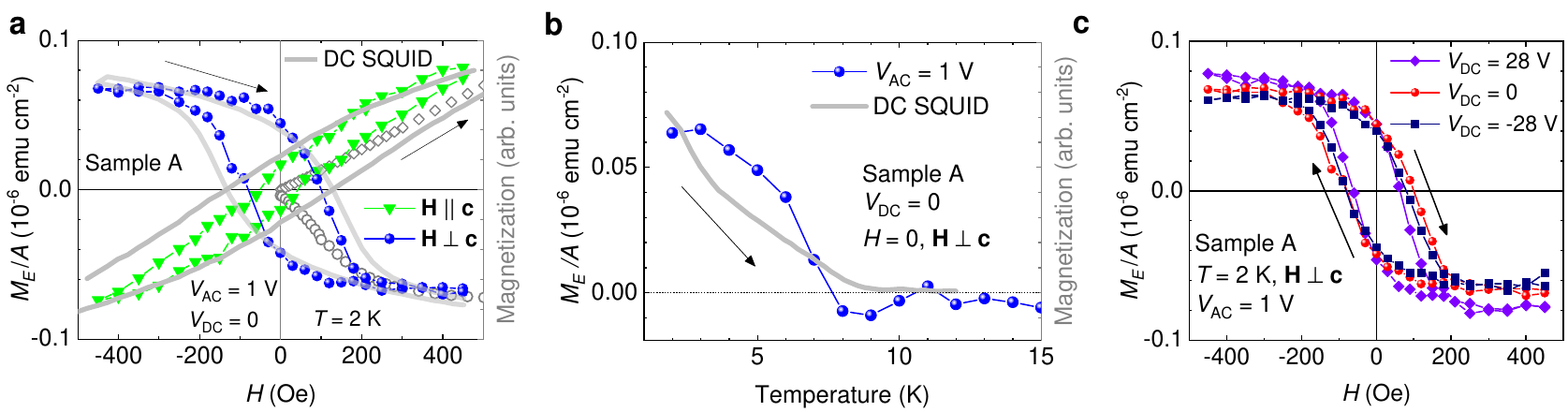}
\caption{\label{Fig:PEME_zoom} {\bf Magnetoelectric effect in the hysteretic regime.} {\bf a}, Areal density of the in-plane magnetic moment amplitude $M_E/A$ of Ga$_{1-x}$Mn$_{x}$N layer with $x=5.6\%$ induced by AC voltage $V_{\text{AC}} = 1$\,V at $T = 2$\,K. The measurements are taken in the in-plane ($\textbf{H} \bot \textbf{c}$) and out-of-plane ($\textbf{H} \| \textbf{c}$) configuration (full circles and triangles, respectively). These results are compared to directly measured magnetizations $-M_{\bot}(H)$ and $M_{\|}(H)$ (grey lines, relative units), measured for Sample B before processing. The virgin magnetization data obtained after cooling in $H \approx 0.15$\,Oe are shown by empty symbols. 
{\bf b}, Areal density of the in-plane magnetoelectric signal $M_E/A$ at $H= 0$ as a function of temperature $T$ (blue circles), compared  also to experimental value $M(T)$ (grey line, relative units). {\bf c}, Magnetoelectric response to $V_{\text{AC}} = 1$\,V measured in the presence of the DC bias $V_{\text{DC}} = \pm 28$\,V applied in the magnetic field  beyond the hysteretic regime.}
\end{figure*}

According to the data displayed in Figs~\ref{Fig:MEsimtheory} and \ref{Fig:PEME_zoom} for Sample A at 2\,K, the electric field decreases magnetization for the magnetic field ${\textbf{H}} \bot {\textbf{c}}$, but increases $M$ for ${\textbf{H}} \| {\textbf{c}}$.  Interestingly, comparing $M_E(H)$ in Fig.~\ref{Fig:MEsimtheory} to $M(H)$ in Fig.~\ref{Fig:Magnetization}, we see that $M_{E}$ is roughly proportional to $\partial M/\partial H$ in high magnetic fields but according to Fig.~\ref{Fig:PEME_zoom}, $M_{E}$ becomes proportional to $M$ in weak magnetic fields. The presented data implies that the converse magneto-electric coupling coefficient, obtained from the relation $\alpha_C =4\pi M_E/E$, attains $7.3\,$~Oe\,cm/MV for wz-Ga$_{0.94}$Mn$_{0.06}$N at 2\,K and in 0.5\,kOe, as $E_{\text{(Ga,Mn)N}} = 0.095$~MV/cm for 0.5\,V applied to Sample A. The highest value of $M_E$ we have detected is $0.23$\,emu/cm$^{3}$ at 5\,V, to be compared to the magnetization saturation value $M_{\text{sat}} = 80$\,emu/cm$^{3}$.

We describe $M_E(H)$ in this easy-plane piezoelectric ferromagnet in terms of an effective anisotropy magnetic field $H_E$ affected by the electric field.  If $H_E \ll H$,
\begin{equation}
M_E(H_j) = M_j(H_j+H_E) - M_j(H_j)\approx \chi_{jj}(H_j)H_E,
\label{eq:ME}
\end{equation}
where $j$ refers to the two explored configurations of the magnetic field; $\chi_{jj} =\partial M_j/\partial H_j$ and, since uniaxial anisotropy is proportional to $\xi(E)$,
\begin{equation}
H_E(H_j) = \pm bd_{33}EM_j(H_j)c/(a\xi),
\end{equation}
where $\pm$ corresponds to ${\textbf{H}} \| {\textbf{c}}$ and ${\textbf{H}} \bot {\textbf{c}}$, respectively, and $b$ is a dimensionless fitting parameter. The above equations define the form of the magnetoelectric coefficient\cite{Fiebig:2005_JAP}  $\alpha_C = M_E/E$ for the case under consideration.

We see in Fig.~\ref{Fig:MEsimtheory} that our macrospin model with experimental values of $\chi_{jj}(H_j)$ and $b = 0.04$ describes quite satisfactorily the field dependence for the two orientations of the magnetic field down to 4\,kOe at $V_{\text{AC}} =0.5$\,V.

While in high magnetic fields $M_E(H_j)\propto \chi_{jj}(H)$, in weak magnetic fields, $M_E(H_j)$ becomes proportional to $M_j(H_j)$, as shown in Fig.~\ref{Fig:PEME_zoom}. This observation supports experimentally  the expectation that $\alpha_C$ is non-zero only under time-reversal symmetry breaking\cite{Fiebig:2005_JAP}, i.e., $\alpha_C$ is odd in $H_j$ in the paramagnetic case, as observed previously\cite{Sztenkiel:2016_NatComm}, or odd in $M_j$ at $H_j\rightarrow 0$, as found here for the ferromagnetic ordering. This means that for sufficiently small values of $M_j$, $M_E$ should be linear in $M_j$, the relation fulfilled by our data displayed in Fig.~\ref{Fig:PEME_zoom}a,b as a function of the magnetic field at 2\,K and as a function of temperature at $H = 0$, respectively.  Accordingly, as shown in Fig.~\ref{Fig:PEME_zoom}b, the magneto-electric effect vanishes near $T_{\text{C}}$, about 7.5\,K for Sample A. Quantitatively, in weak magnetic fields, $M_E(H_j) = \pm \beta M(H_j)$, where $\beta =2\times10^{-3}$  for Sample A and $V_{\text{AC}} =1$\,V.

As presented in Fig.~\ref{Fig:PEME_zoom}c,  we have also checked that the application of the DC bias of either polarization has no effect on $M_E(H)$ generated by $V_{\text{AC}}$. This  finding is consistent with the fact that the DC electric field is non-zero only across the HfO$_2$ film.

As far as the origin and magnitude of the parameters $b$ and $\beta$ are concerned, we note that, in general, crystalline magnetic anisotropy, which we perturb by the electric field $via$ the inverse piezoelectric effect, originates from a single-ion contribution and anisotropy of spin-spin coupling. In magnetic compounds in question, the spin-orbit interaction within anions transmitting exchange between localized spins plays a dominant role\cite{Larson:1988_PRB,Dietl:2001_PRB}, so that this anisotropy should be relatively small for the light nitrogen ions. This fact suggests that single-ion anisotropy, rather sizable for Mn$^{3+}$ ions with the orbital momentum $L = 2$, dominates in (Ga,Mn)N.  The  signs of $b$ and $\beta$ are consistent with this expectation. However, though we know rather precisely the crystal field parameters of Mn$^{3+}$ ions in GaN and their dependence on $\xi$ (ref.~\onlinecite{Sztenkiel:2016_NatComm}), there are persistent difficulties in relating that information with parameters of the macrospin model\cite{Frobrich:2006_PR}, particularly in dilute magnetic systems\cite{Sawicki:2018_PRB} and in the percolating case. This fact makes a theoretical determination of $b$ and $\beta$ difficult within the macrospin model. We have, therefore, decided to adapt a stochastic Landau-Lifshitz-Gilbert (LLG) atomistic spin model\cite{Skubic:2008_JPhysCM,Evans:2014_JPhysCM, Evans:2015_PRB} for the case under consideration.

 \ \\
{\bf Landau-Lifshitz-Gilbert atomistic spin model}.  The way that approach\cite{Evans:2014_JPhysCM, Evans:2015_PRB} is adopted to ferromagnetic (Ga,Mn)N is detailed in Methods. Our atomistic LLG model was also employed to describe magnetization of (Ga,Mn)N in the paramagnetic regime, where magnetic properties are determined by single Mn ions, nearest neighbour Mn pairs, and triangles, so that the outcome of the LLG model which treats the Mn spin classically, could be compared to the results of computations describing the spins quantummechanically\cite{Edathumkandy:2022_JMMM}.

The magnetic energy of Mn spins is described by a spin Hamiltonian $\mathcal{H}$ taking into account the Zeeman energy, Heisenberg spin-spin interactions between Mn spins located up to the 18th nearest neighbour positions, and the magnetocrystalline anisotropy energy,  $\mathcal{H}=\mathcal{H}_{\mathrm{Z}}+\mathcal{H}_{\mathrm{Exch}}+\mathcal{H}_{\mathrm{Anizo}}$,  where $\mathcal{H}_{\mathrm{Anizo}}$ includes the Jahn-Teller contribution and the trigonal anisotropy term taken as,
\begin{equation}
\label{eq:HTrig}
\mathcal{H}_{\mathrm{Trig}}=-\frac{1}{2}K_{\mathrm{Trig}}{\sum_{i}}\frac{1}{2}[{S}_{iz}^2-( {S}_{ix}^2+{S}_{iy}^2 )].
\end{equation}
Here, $K_{\mathrm{Trig}}\propto \xi$ is the anisotropy parameter describing the trigonal deformation that can be controlled by the electric field $E$, and $\textbf{S}_i$ is a unit vector describing the local magnetic moment direction at the site $i$.  The explicit forms of $\mathcal{H}_{\mathrm{JT}}$ and $\mathcal{H}_{\mathrm{Exch}}$ together with the values of relevant exchange integrals, are given in Methods.

The results on thermal equilibrium values of macroscopic magnetizations $M_j(H_j)$ without and with an electric field $E$ and for two directions $j$ of the external magnetic field are obtained for a  simulation box $21 \times 21 \times 18$~nm$^3$ containing 169000 cation sites, 9464 ($5.6\%$) of which encompass randomly distributed spin vectors. The site number is limited only by the available hardware employed to trace the time evolution towards thermal equilibrium values $M_j(H_j)$ over an acceptable time frame. The simulations start in 50\,kOe, in which Mn spins are aligned along the magnetic field, and next the field is decreased in steps specified in Methods.


Considering the lack of a clear correspondence between parameters of the quantum and classical spin Hamiltonians, our strategy involves two steps.
First, by fitting the LLG model to the experimental data on $M_j$ vs.~$H_j$ at $E = 0$ and for the two orientations of the external magnetic field, as presented in Fig.~\ref{Fig:Magnetization}, we have evaluated $K_{\mathrm{JT}}=0.85$~meV and $K_{\mathrm{Trig}}=0.1$~meV appearing in system Hamiltonian.
A positive sign of $K_{\mathrm{Trig}}$ is in agreement with the data on $\xi$ obtained from x-ray diffraction (XRD) measurements, shown in Fig.~\ref{Fig:SamplePiezo}a. However, it should be noted that the simulations point to  larger remanence than observed experimentally.  Moreover, even rescaling Curie temperature obtained from the numerical modeling for classical spins by a factor $(S+1)/S$, we obtain $T_\mathrm{C}^{\text{sim}}\cong 4.5$~K, which differs significantly from the experimental value $T_\mathrm{C} \cong 7.5$~K.

\begin{figure*}[htb]
\centering
\includegraphics[width=17 cm]{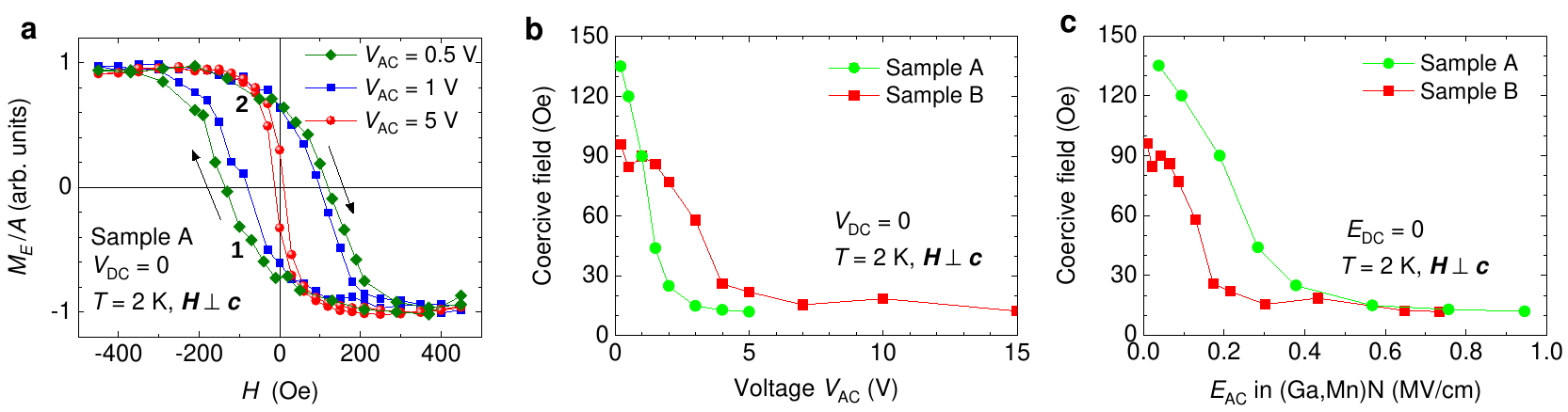}
\caption{\label{Fig:PEME_coercivity} {\bf Effect of the AC voltage on coercivity.}  {\bf a}, Areal density of the in-plane magnetic moment amplitude $M_E/A$ generated by three different values of the AC voltage $V_{\text{AC}}$ as a function of the in-plane magnetic field.  {\bf b, c}  Corresponding coercivity  field as a function of $V_{\text{AC}}$ and the electric field $E_{\text{AC}}$ across the (Ga,Mn)N layers of thicknesses 10 and 160~nm for Samples A (green circles) and B (red squares), respectively.}
\end{figure*}

Second, we have performed computations for the determined
value of $K_{\mathrm{JT}}$ and for $K_{\mathrm{Trig}}$ in the absence
and presence of the electric field,
\begin{equation}
K_{\mathrm{Trig}}(E) = K_{\mathrm{Trig}}(0)[1 + d_{33}Ec/(a\xi)],
\end{equation}
where $\xi = 0.0029(3)$ for Sample A according to Fig.~\ref{Fig:SamplePiezo}a.
Here, in order to reduce uncertainties resulting from statistical fluctuations of computational results,  we considered $E$ corresponding to $V_{AC} = 5$\,V, and reduce the numerical values of $M_E(H_j)$ by a factor of ten to compare to experimental results depicted in Fig.~\ref{Fig:MEsimtheory}  for $V_{AC} = 0.5$\,V. As seen in Fig.~\ref{Fig:MEsimtheory}, the LLG theory is successful only qualitatively, i.e., predicts properly the sign and shape of $M_E(H_j)$.  The obtained results bring us to the conclusion that the LLG approach, when applied to dilute ferromagnets, predicts too large magnitudes of remanence in the case of both $M(H_j)$ and $M_E(H_j)$. We assign this outcome to time scale of our experiment, which is sufficiently long to allow demagnetization of most Mn spins outside the percolation cluster. Such a demagnetization process does not occur in the necessary much shorter time frame of the LLG simulations, so that most of Mn spins contribute to the theoretical remanence, as shown by solid lines in Fig.~\ref{Fig:Magnetization}.

%
\ \\
{\bf{Nonlinearities and irreversibilities}}. We now discuss surprising switching effects found in the hysteretic regime. The magnetoelectric signal $M_E(H)$ in this range and at 2\,K is plotted in Fig.~\ref{Fig:PEME_coercivity}a as a function of the in-plane magnetic field ${\textbf{H}}\bot {\textbf{c}}$ for  three different $AC$ modulation voltages $V_{\text{AC}}$. As seen, the width of the coercive field $H_c$ gets sizably reduced (from 120~Oe to about 15~Oe) by  increasing $V_{\text{AC}}$ to 5\,V. The effect is better visualized in Fig.~\ref{Fig:PEME_coercivity}b,c, where we present changes of the coercive field with the applied voltage $V_{\text{AC}}$ to the whole structure and with the corresponding electric field $E_{\text{AC}}$ across the (Ga,Mn)N layer of the thickness $d = 10$ and 160\,nm for Sample A and B, respectively. We see that the $M_E$ appears highly nonlinear in $E_{\text{AC}}$ in the hysteretic region. According to Fig.~\ref{Fig:PEME_coercivity}c, this nonlinearity is stronger in Sample B in which, judging by the magnitude of $H_c$, ferromagnetism is softer than in Sample A. Altogether, these results imply that taping of the system by sufficiently high $V_{\text{AC}}$ brings the system towards thermal equilibrium.

While a time-independent DC bias has no effect in our structures (see Fig.~\ref{Fig:PEME_zoom}c), we find that step changes of the DC voltage, if applied
in the non-equilibrium range of the hysteresis, moves magnetization toward the direction of the external magnetic field. Our experiment begins by  sweeping  the external magnetic field to the non-equilibrium range of the hysteresis, and then we change $V_{\mathrm{DC}}$  in a step-like manner by $\Delta V_{\text{DC}}$ at constant $H$ monitoring $M_E$ at selected voltage values $V_{\mathrm{DC}}=$0, 3,..., 18~V or, for the negative sign of $\Delta V_{\text{DC}}$, at $V_{\mathrm{DC}}=$0, -3, ..., -18~V. There is at least 1\,sec delay between consecutive voltage changes.

\begin{figure*}[htb]
\centering
\includegraphics[width=15.0 cm]{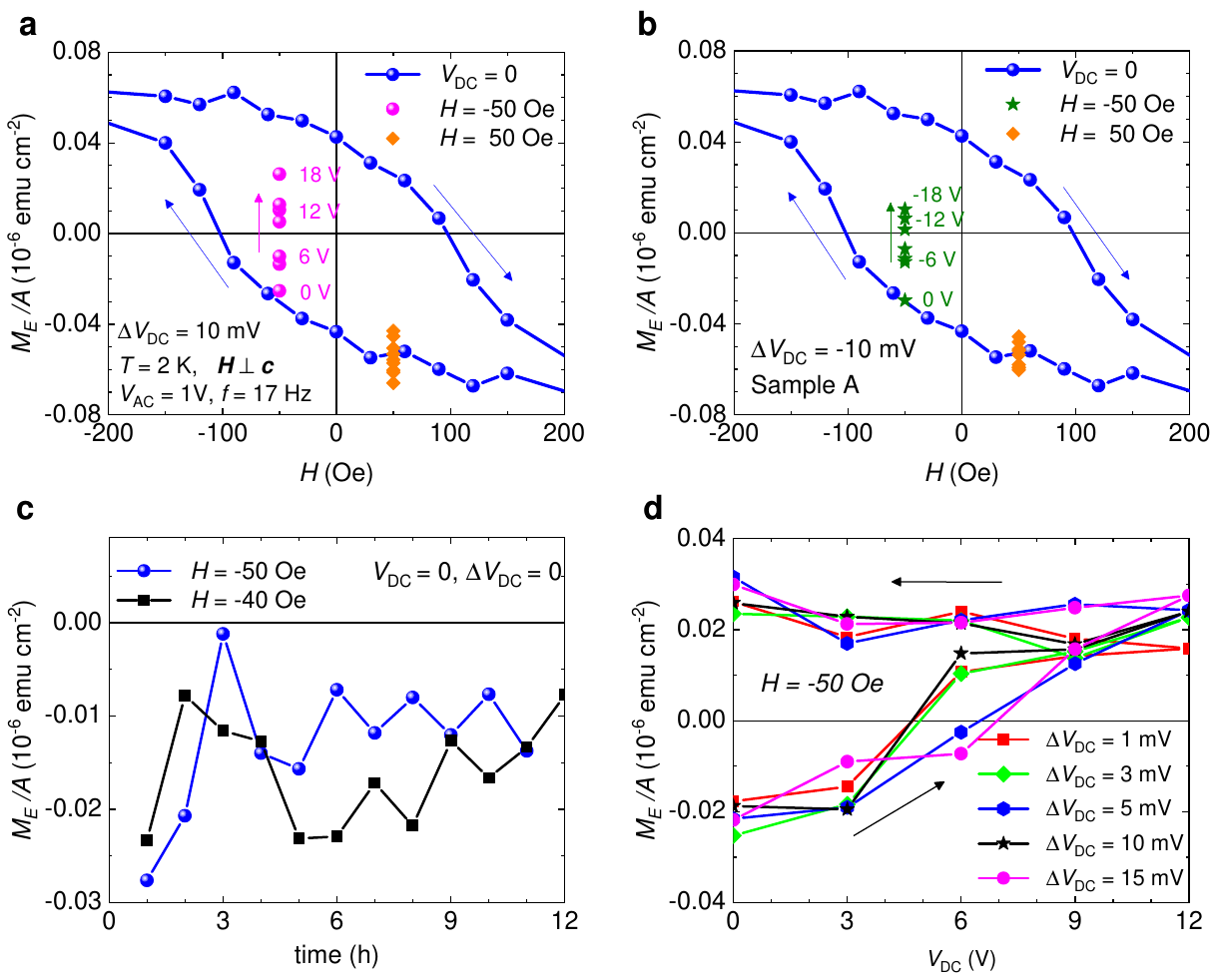}
\caption{\label{Fig:Switching} {\bf Magnetization switching by DC voltage changes}. {\bf a,b}, At $V_{\mathrm{DC}}=0$\,V, the magnetic field is swept from -450 to +50\,Oe   or to -50\,Oe (blue circles, full sweep is shown). The $V_{\mathrm{DC}}$ is then changed in step-like manner by $\Delta V_{\text{DC}}= 10$\,mV or -10\,meV, and the resulting magnetoelectric signal $M_E$ is measured at successive points $V_{\mathrm{DC}}=0$, 3,..., 18\,V (orange points or magenta points in {\bf a}) or $V_{\mathrm{DC}}=0$, -3, ..., -18\,V (orange points or green points in {\bf b}). Irreversible switching of magnetization is observed independently of the DC voltage polarization if for $H<0$, $M>0$ (i.e., $M_E <0$ for ${\textbf{H}}\bot {\textbf c} $ configuration), meaning that the system is initially out of thermal equilibrium. {\bf c} Magnetoelectric response $M_E$ as a function of time for $\Delta V_{\mathrm{DC}}=0$. {\bf d}, Magnetoelectric response $M_E$ as a function of $V_{\mathrm{DC}}$ for various magnitudes of DC voltage steps $\Delta V_{\mathrm{DC}}$.}
\end{figure*}

Results of such investigations are collected in Fig.~\ref{Fig:Switching} which shows that:  (i) Magnetization changes generated by voltage alterations occur only in the non-equilibrium regime and those changes are irreversible, i.e. magnetization does not change under removal of voltage. (ii) According to results depicted in Figs~\ref{Fig:Switching}a,b, outside the metastable regime, the observed drift in $M_E$ magnitudes for consecutive voltage changes is of the order of time-dependent fluctuations in the $M_E$ values, shown in Fig.~\ref{Fig:Switching}c. (iii) Voltage-induced shifts in $M_E(V_{\mathrm{DC}})$ values are independent of the DC voltage polarization, and a sufficiently high magnitude of $|V_{\mathrm{DC}}|$ reverses the magnetization direction (Fig.~\ref{Fig:Switching}a,b). (iv) As shown in Fig.~\ref{Fig:Switching}d, the consecutive values of $M_E(V_{\mathrm{DC}})$ are also virtually independent of the number of steps leading to a given $V_{\mathrm{DC}}$, at least in the range of step heights $1 \leq \Delta V_{\text{DC}} \leq 15$\,meV.

It is remarkable that irreversible shifts of magnetization toward the thermal equilibrium value by voltage alterations, as we report here, were also detected at 0.25\,K by local SQUID magnetometry in gated structures of another dilute ferromagnetic semiconductor (Bi,Sb,Cr)$_2$Te$_3$ (ref.~\onlinecite{Lachman:2015_SA}). As in our case, the effect was independent of gate voltage polarization and the magnetization alteration scaled with the magnitudes of the gate voltage excursion. Also, time-independent gate voltage had little effect on magnetization and hystereses. Furthermore, spatially-resolved magnetization studies pointed to the presence of superparamagnetic regions with noticeable magnetization relaxation over the laboratory time scale.

To explain the origin of that universal behaviour, we first recall that despite magnetic dilution, a well-defined magnetic domain structure was found in (Ga,Mn)As, the fact explained quantitatively in terms of a long-range character of spin-spin interactions in that type of ferromagnets\cite{Dietl:2001B_PRB}. In order to address this question in (Ga,Mn)N, we show in Fig.\,4a experimental data on virgin magnetization obtained after cooling the sample down to 2\,K under near-zero-field conditions ($H \approx 0.15$\,Oe). The measured form of $m(H)$ corresponds neither to the domain-wall motion nor to the domain-wall pinning in a typical multidomain case for a ferromagnet. In the former case one would expect a rapid increase of m, nearly along the $y$-axis, in the latter the initial m(H) would be horizontal along the $x$-axis. Actually, the observed behaviour is consistent with or even supports the view that in the case of short-range interactions, driven here by superexchange, the virgin magnetic properties are dominated more by superparamagnetic regions present in the percolation picture than by a regular domain structure which is formed by dipole interactions in dilute ferromagnets with long-rage exchange couplings\cite{Dietl:2001B_PRB}. 

More specifically, we expect in both (Bi,Sb,Cr)$_2$Te$_3$ and (Ga,Mn)N, a co-existence,  at $T_{\text{C}} >  T > 0$, of an infinite ferromagnetic percolating clusters with superparamagnetic regions, characterized by a broad spectrum of blocking temperatures and relaxation times. With no doubts, the presence of long-range dipole interactions, responsible for the domain formation in materials like (Ga,Mn)As, contributes to the complexity of spin dynamics. Therefore,  (Bi,Sb,Cr)$_2$Te$_3$ and (Ga,Mn)N in the metastable regime can be regarded as a glass that is far from thermal equilibrium. A question arises: how do such media respond to a time-dependent perturbation? It has been theoretically demonstrated that in driven glassy or granular systems, the degrees of freedom governing slow relaxation thermalize to an effective temperature, whose deviation from bath temperature scales with the drive strength\cite{Makse:2002_N,Loi:2008_PRE,Berthier:2013_NP}. Considering a steady decay of the coercive force with temperature, we see that the data on dilute ferromagnetic semiconductors with a short range of spin-spin interactions provide the experimental corroboration of theoretical predictions concerning tapped glassy systems. However, neither in (Ga,Mn)N, nor in (Bi,Sb,Cr)$_2$Te$_3$ a full magnetization reversal has been attained by
time-dependent voltage manipulations, at least in the explored time-domain much longer than the inverse precession frequency. This fact is in accord with our scenario that implies the presence of clusters in which a ferromagnetic spin arrangement is robust to weak perturbations.

To the best of our knowledge, the experiments have revealed novel aspects of driven dynamics, such as the fact that the trajectory toward the thermal equilibrium depends only on the accumulated magnitude of the drive changes but not on the number of those changes. It is worth noting that the standard Langevin dynamics does not capture the effects of time-dependent perturbations, so that the nonlinear and irreversible effects do not show up in our LLG simulations.

\section*{Conclusions}
In our work, we have exploited the recent progress in the MBE growth of high quality quantum structures incorporating ferromagnetic (Ga,Mn)N to examine much needed spintronic functionality, i.e., the manipulation of magnetization by an {\em electric} field. Our results demonstrate that (Ga,Mn)N with a sufficiently large Mn concentration is one of rather rare piezoelectric ferromagnetic homogeneous compounds, in which the magnetoelectric effect can be detected and, moreover, examined theoretically in a quantitative manner. In another perspective, the magnetization manipulation enlarges a pallet of wide-ranging nitride functionalities\cite{Jena:2019_JJAP} that have made GaN and related systems the second most important semiconductor family next to Si. In particular, GaN:Mn serves already as a semiinsulating substrate\cite{Bockowski:2018_JCG} in device epitaxy. The spectrum of (Ga,Mn)N functionalities has recently been  enriched by the demonstration of efficient
generation and detection of spin currents in Pt/(Ga,Mn)N structures up to 50\,K (ref.\,\onlinecite{Mendoza-Rotarde:2024_APL}) using spin Hall magnetoresistance (SMR) effect\cite{Nakayama:2013_PRL}. The same SMR could be then employed to probe the magnetic state of the insulating (Ga,Mn)N in the electrical way, paving the way towards more practical applications. 

Our experimental results have revealed the presence of two distinct regimes showing a different pool of phenomena. The first of them concerns a wide range of magnetic and electric fields, in which the magnetoelectric response $M_E$ is linear in the AC electric field $E$ and reversible, i.e., the magnetization magnitude returns to its initial value after switching off the electric field. We have developed for that case (i) the macrospin approach and (ii) the atomistic theory with Landau-Lifshitz-Gilbert equations containing the external magnetic field, molecular fields generated by neighbouring Mn spins, and Gaussian magnetic noise -- the Langevin dynamics. The macrospin model, containing one fitting parameter, describes $M_E$ for both magnetic field configurations, ${\textbf{H}}\bot {\textbf{c}}$ and ${\textbf{H}}\| {\textbf{c}}$ over a wide field range, including the hysteretic regime. In the case of the LLG theory, isotropic ferromagnetic Mn-Mn exchange integrals come from the theory which describes $T_{\text{C}}$(x), whereas parameters of the single-ion anisotropy, from fitting $M(H)$ for the two configurations. A general agreement between experimental and theoretical results on $M_E(H)$  demonstrates that the magnetoelectric response results primarily from the influence of the inverse piezoelectric effect on the wurtzite $c$ lattice parameter, which affects the trigonal uniaxial single-ion magnetocrystalline anisotropy. At the same time, we have identified a limited predictive power of the LLG simulations concerning the magnitudes of remanent magnetization and magnetoelectric effect in dilute ferromagnets at $H \rightarrow 0$.

The other magnetoelectric phenomena discussed in our work appear in the metastable range of the low-field hystereses. Our results, and also those found for (Bi,Sb,Cr)$_2$Te$_3$\cite{Lachman:2015_SA}, let us to suggest that dilute ferromagnetic semiconductors with short-range spin-spin couplings show, in the metastable regime, characteristics similar to glasses far from thermal equilibrium. One of pertinent questions is the response of such systems to time-dependent perturbations specified by a time scale long compared to both the inverse Larmor frequency and the transverse spin relaxation time $T_2$, governed by non-scalar spin-spin interactions, but comparable to $T_1$ of superparamagnetic clusters incorporating most of localized spins at $T>0$. To explain characteristics of irreversible magnetoelectric responses in (Ga,Mn)N and (Bi,Sb,Cr)$_2$Te$_3$, we refer to extensive numerical simulations of glasses\cite{Makse:2002_N,Loi:2008_PRE,Berthier:2013_NP}, indicating that time-dependent perturbations of such systems lead to fast thermalization to an enhanced temperature magnitude facilitating a unidirectional drift of the magnetization value and direction toward the thermal equilibrium state. Conversely, magnetoelectric data for relatively simple and easy controllable systems, such as
(Ga,Mn)N and (Bi,Sb,Cr)$_2$Te$_3$, can motivate further theoretical studies of non-equilibrium physics in disordered systems.

\section*{Methods}

\subsection*{Material}
\begin{figure}[tb]
\centering
\includegraphics[width=8.5 cm]{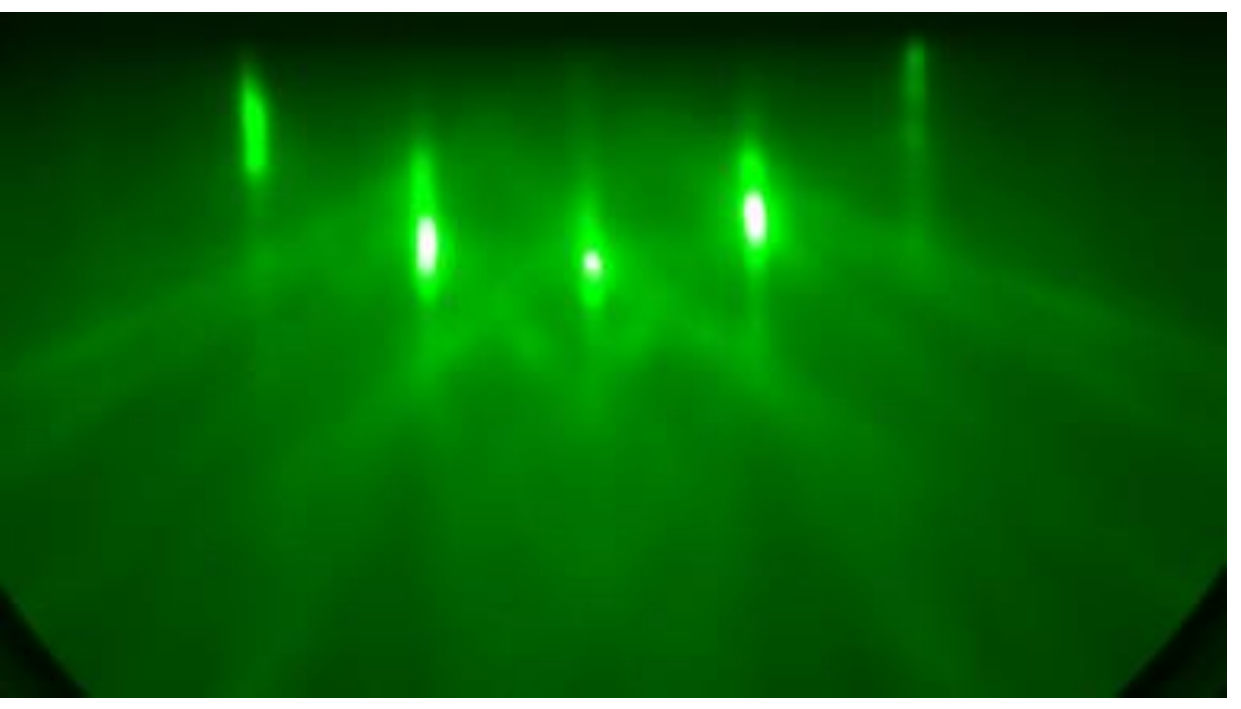}
\caption{\label{Fig:M_RHEED_M20043} {\bf Reflection high-energy electron diffraction (RHEED).} RHEED pattern along [11$\bar{2}$0] azimuth of Sample B observed after the growth and cooling the sample down to 200$^o$C. }
\end{figure}

The (Ga,Mn)N layers investigated here have been grown by plasma assisted molecular beam epitaxy (MBE) method in a Scienta-Omicron Pro-100 MBE at 620$^{\circ}$C under near stoichiometric conditions following closely the previously elaborated protocol\cite{Gas:2018_JALCOM}.
In order to reduce the number of threading dislocations and simultaneously provide the back electrical contact to the final capacitive device we use commercially available conductive low threading dislocation density (LTDD) $n$-type (0001) GaN:Si templates deposited on a $c$-plane sapphire, which contain, generally termed, an SiN$_x$ mask.
During the growth the surface quality of the layers has been investigated \textsl{in situ} by reflection high-energy electron diffraction (RHEED).
The post growth [11$\bar{2}$0] RHEED pattern of Sample B collected after the growth (at a temperature below 200$^o$C) is shown in Fig.~\ref{Fig:M_RHEED_M20043}.
The streaky and sharp RHEED pattern  with clearly visible Kikuchi lines indicates that the (Ga,Mn)N surface is
smooth and relatively free of oxides or other contaminants.

Two samples have been prepared for this study.
Sample A of a low thickness of (Ga,Mn)N layer, $d_{\mathrm{(Ga,Mn)N}}=10$~nm  and a thicker Sample B with $d_{\mathrm{(Ga,Mn)N}}=160$~nm.
The precise values of $d_{\mathrm{(Ga,Mn)N}}$ are established upon high resolution x-ray characterization described below.
Both are having a similar Mn concentration $x$, about 6\%, established by direct magnetic measurements.

\subsection*{Structural characterization}
\begin{figure*}[th]
\centering
\includegraphics[width=16 cm]{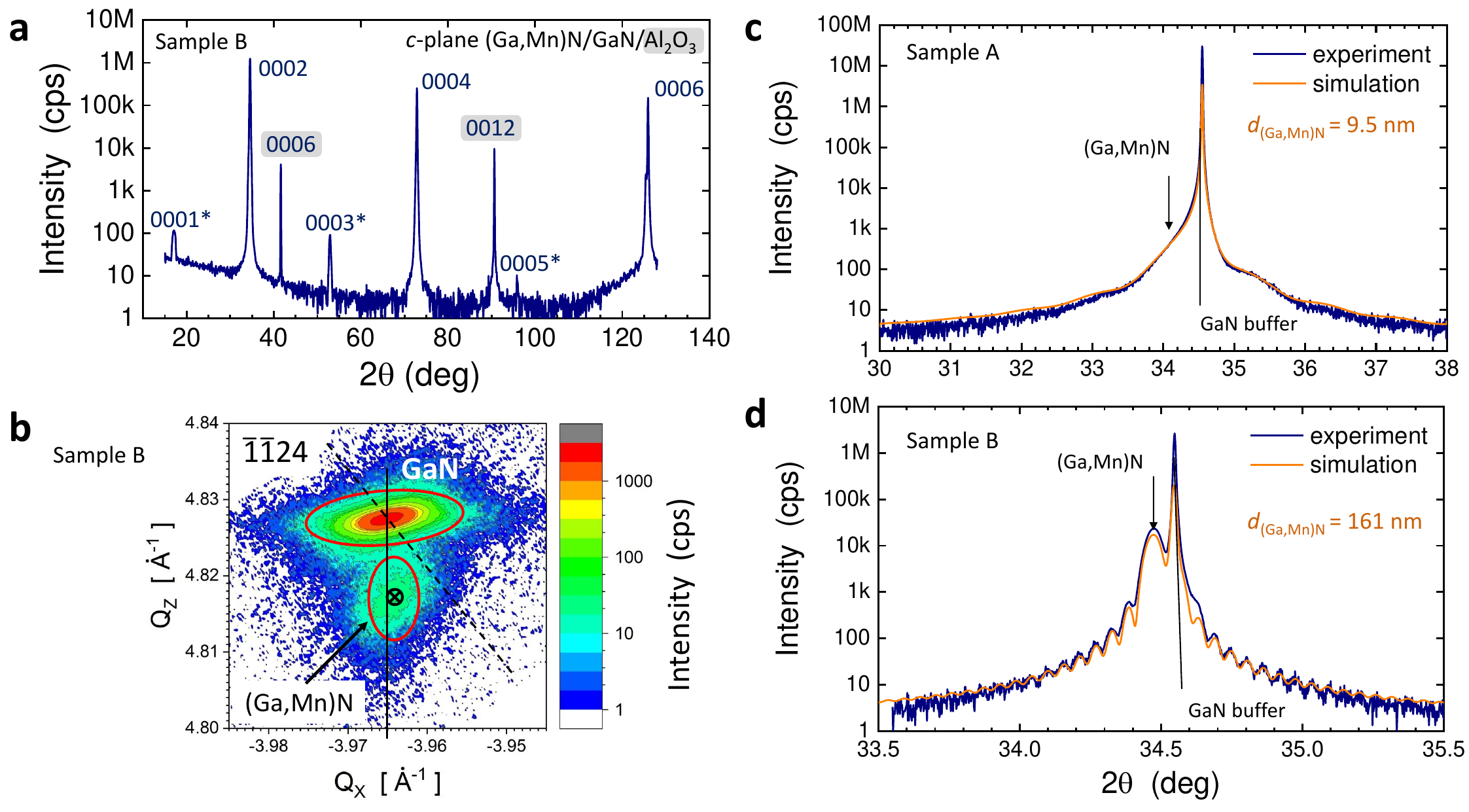}
\caption{\label{Fig:M_XRD_M20043_All}  {\bf High-resolution (HR) x-ray characterization of investigated samples.}  {\bf a} Diffraction pattern of Sample B in $2\theta / \omega$ mode. Three GaN reflections denoted with a star are a result of a multiple reflection and are seen here due to a specific orientation of the sample. The reflections indexed on the grey background are from the sapphire substrate.  {\bf b} The reciprocal space map of the asymmetric $\bar{1}\bar{1}$24 reflection of Sample B. The high intensity node is related to the GaN template layer. The low intensity node reflects a practically fully strained (Ga,Mn)N layer, as can be inferred from the location of this node very close to the  vertical line, which, together with the inclined dashed line define the triangle of relaxation.
{\bf c, d} HR diffraction pattern for the 0002 Bragg reflection for Sample A and B, respectively (the dark blue lines). The orange lines marks results of simulations upon which the values of the layers thicknesses are established (given in the legends).}
\end{figure*}
The crystal structure, epitaxial relationship, in- and out-of-plane lattice parameters, and the strain condition of the samples are established by high resolution x-ray diffraction (HRXRD) using a Philips  X'Pert MRD diffractometer operating on wavelength 1.5406~\AA of Cu$_{K\alpha 1}$ line.
Its incident beam is formed by an x-ray mirror and a monochromator [four-bounce (220) Ge asymmetric cut].
Figure~\ref{Fig:M_XRD_M20043_All} collects the most relevant findings.
Panel {\bf a} demonstrates a $2\theta/\omega$ scan measured for Sample B over a wide range of detector angle changes, indicating that within a dynamic range of intensity exceeding four orders of magnitude no reflections specific to inclusions of other phases are seen.
The reciprocal space map of the $\bar{1}\bar{1}$24 reflection is given in panel {\bf b}.
Two main features can be recognized there: the high intensity peak located at $Q_z = 4.827$~\AA$^{-1}$ - related to the GaN template, and a much weaker one at smaller reciprocal lattice units, at $Q_z = 4.817$~\AA$^{-1}$, from the thinner (Ga,Mn)N layer.
The map indicates  a nearly perfect pseudomorphic growth, meaning that the (Ga,Mn)N layer adopts the same in-plane lattice parameter as that of the substrate (the GaN LTDD template).

HRXRD patterns for the 0002 Bragg reflection are marked as the thicker blue line in Figs.~\ref{Fig:M_XRD_M20043_All}c,d for Samples  A and B, respectively.
Clearly resolved x-ray interference fringes  imply a high structural perfection of the layers and good quality of the interfaces.
The peaks corresponding to the (Ga,Mn)N epitaxial layers are shifted to smaller angles with respect to that of the GaN buffer, a result of a larger perpendicular lattice parameter of (Ga,Mn)N.
The orange lines mark results of simulations performed in the frame of the dynamical theory of x-ray diffraction using commercially available PANalytical EPITAXY software.
In these computations we put the elastic stiffness constants in (Ga,Mn)N as those in bulk GaN 
and $c=5.1850(5)$~\AA \, and $a=3.1885(5)$~\AA, as established in ref.~\onlinecite{Leszczynski:1999_JALCOM}.
The best fit to the experimental data is obtained for $d_{\mathrm{(Ga,Mn)N}}=9.5(1)$~nm for Sample A and $d_{\mathrm{(Ga,Mn)N}}=161(1)$~nm for Sample B, from which simplified values of 10 and 160\,nm values are adopted to characterize both layers.
We can parenthetically add that allowing for 5\% of lattice relaxation (to account for a small horizontal shift of the (Ga,Mn)N node seen in Fig.~\ref{Fig:M_XRD_M20043_All}b) does not call for a greater modification of $d_{\mathrm{(Ga,Mn)N}}$  than the modeling uncertainty specified above.

The XRD measurements are also used to establish the values of the trigonal distortion parameter $\xi = c/a-\sqrt{8/3}$.
Generally, the value of $c$ is obtain from the symmetrical 0002 reflection,
$a$ parameter is calculated based on data from both 0002 one and asymmetrical -1-124 reflections.
Bragg angle of $\bar{1}\bar{1}$24 is usually determined from a relevant reciprocal space map, as the one presented in Fig.~\ref{Fig:M_XRD_M20043_All}b.
The bare results for $\xi$ obtained from XRD in Samples A and B are: $\xi_A = 0.0029(3)$ and $\xi_B = -0.0005(3)$, respectively.
However, these values quantify the distortion at room temperature, whereas PEME measurements are performed at 2\,K.
So one can expect an additional deformation in response to the specific thermal contractions of all the constituent layers embedded within the structure.
In order to assess the magnitude of the temperature induced modification of the trigonal distortion of the (Ga,Mn)N unit cell upon cooling we  transform $\xi(300$~K) into $\xi(4$~K) by taking into account\cite{Cullity:2001_book,Kidd:2015_booklet} relevant elastic constants\cite{Adachi:2016_JAP} and thermal expansion coefficients of $a$ and $c$ in sapphire\cite{Lucht:2003_JACr} and GaN, ref.~\onlinecite{Kirchner:2000_APL}.
We find that for this particular material combination [the layer stack is presented in a later section (Fig.~\ref{Fig:M_sample})].
 the magnitude of the expected change of $\xi$ upon cooling to liquid helium temperatures is marginally small.
The calculated change does not exceed +0.0001, which is only a third of the typical uncertainty of establishing $\xi$ from HRXRD.
We report all these values in Table~\ref{tab:ksis}.

\vspace{0.1cm}
\begin{table}
\caption{\textbf{Parameters of sample A and B.} (Ga,Mn)N layer thicknesses, $d_{\mathrm{(Ga,Mn)N}}$, and magnitudes of trigonal deformation parameter $\xi =c/a-\sqrt{8/3}$ established by high resolution x-ray diffraction at 300 K. The corrected magnitudes of $\xi$ at liquid helium temperatures, $\xi(4$~K), and the typical experimental error, $\delta(\xi)$ are also given.}
\label{tab:ksis}
\begin{ruledtabular}
\begin{tabular}{ccrrrl}
Sample & $d_{\mathrm{(Ga,Mn)N}}$ & $\xi(300$~K) & $\xi(4$~K)& $\delta(\xi)$ & \\
\hline
A   &  10 & 0.0029 & 0.0029 & 0.0003 & \\
B  & 160 & -0.0005 & -0.0004 &0.0003 & \\
\end{tabular}\\
\end{ruledtabular}
\end{table}

Importantly for this project, the atomic force microscopy AFM investigations yield a low surface roughness.
As reported in ref.~\onlinecite{Kalbarczyk:2019_JALCOM} for the same Sample B and in ref.~\onlinecite{Gas:2018_JALCOM} for a range of sistering samples, the root mean square of the roughness of the surface does not exceed 2~nm on a $5 \times 5$~$\mu$m$^2$ field of view.
The main reason of the surface roughening is the formation of spiral hillocks  known to form at the locations where the dislocations piercing the (Ga,Mn)N layer reach the free surface, otherwise clear monoatomic steps are observed.
The hillocks' tops are these locations on the surface of GaN, which provide short-circuiting electrical passes for the vertical current across the layer\cite{Hsu:2001_APL}, so they have to be blocked for achieving effective gating of the material.
It was previously estimated\cite{Kalbarczyk:2019_JALCOM} that the surface density of hillocks in our samples, grown on LTTD substrates, does not exceed $10^8$\,cm$^{-2}$, which represents an improvement exceeding two orders of magnitude compared to typical growth on plain GaN-templated sapphire substrates.

The growth has been performed on quarters of 2" wafers.
In order to avoid a high lateral gradient of the Mn concentration, and so of the other magnetic characteristics  caused by the inhomogeneous substrate temperature during the MBE growth\cite{Gas:2018_JALCOM}, for the current study the specimens are cut from the center of the quarter-wafers substrates, where the homogeneity of Mn concentration and related magnetic characteristics is the best\cite{Gas:2018_JALCOM,Gas:2020_JALCOM}.

\subsection*{Magnetic characterization}

All the magnetic measurements are performed using a Quantum Design MPMS XL superconducting quantum interference device (SQUID) magnetometer following a procedure elaborated earlier\cite{Sztenkiel:2016_NatComm,Sawicki:2011_SST,Gas:2019_MST} and adjusted for measurements of ultra-thin (Ga,Mn)N films\cite{Gas:2021_JALCOM} deposited on sapphire substrates\cite{Gas:2022_Materials}. As shown previously for MBE-grown epilaers\cite{Kunert:2012_APL,Gas:2018_JALCOM,Stefanowicz:2013_PRB}, and checked for the present samples, our high-resolution magnetic data do not reveal the presence of any high-$T_{\text{C}}$ ferromagnetic precipitation.

\subsection*{SQUID magnetoelectric measurements}

\begin{figure}[tbh]
\centering
\includegraphics[width=7.0 cm]{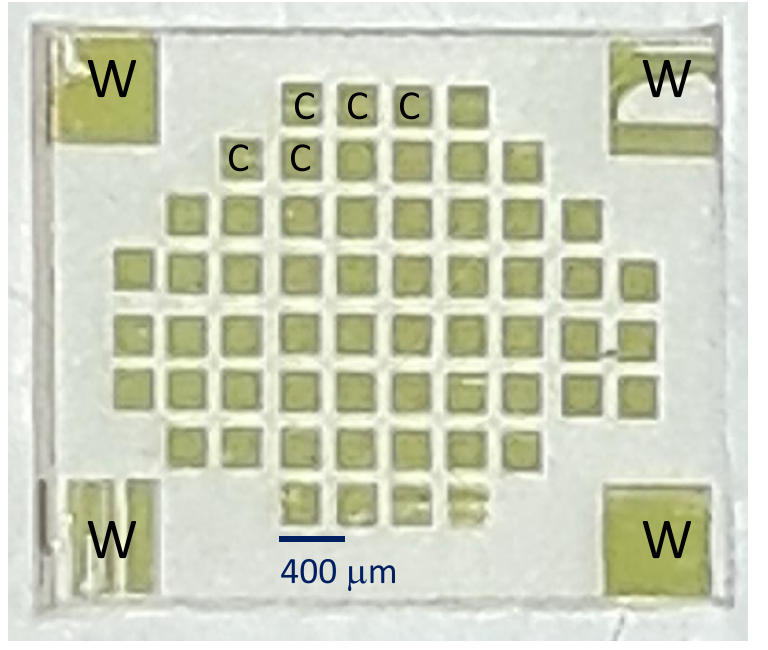}
\caption{\label{Fig:M_sample} {\bf Sample A right after the lithographic processing.} Gold pads W at the corners are the contacts to the back gate layer of $n^+$-GaN:Si, the topmost layer of the substrates on which the 10\,nm-thick (Ga,Mn)N layers have been deposited. Small contact pads C are deposited on 60~nm HfO$_2$-dielectric oxide deposited by atomic layer deposition onto the surface of (Ga,Mn)N to prevent electrically shorts via the threading dislocations. Each pad forms a parallel plate capacitor with the back gate conductive layer of GaN:Si. Upon further electrical testing only the non-leaking small capacitor (up to 12) are connected in parallel to form the final metal-oxide-semiconductor capacitor structure with capacitance of up to 1.2~nF.}
\end{figure}
\begin{figure}[th]
\centering
\includegraphics[width=8.5 cm]{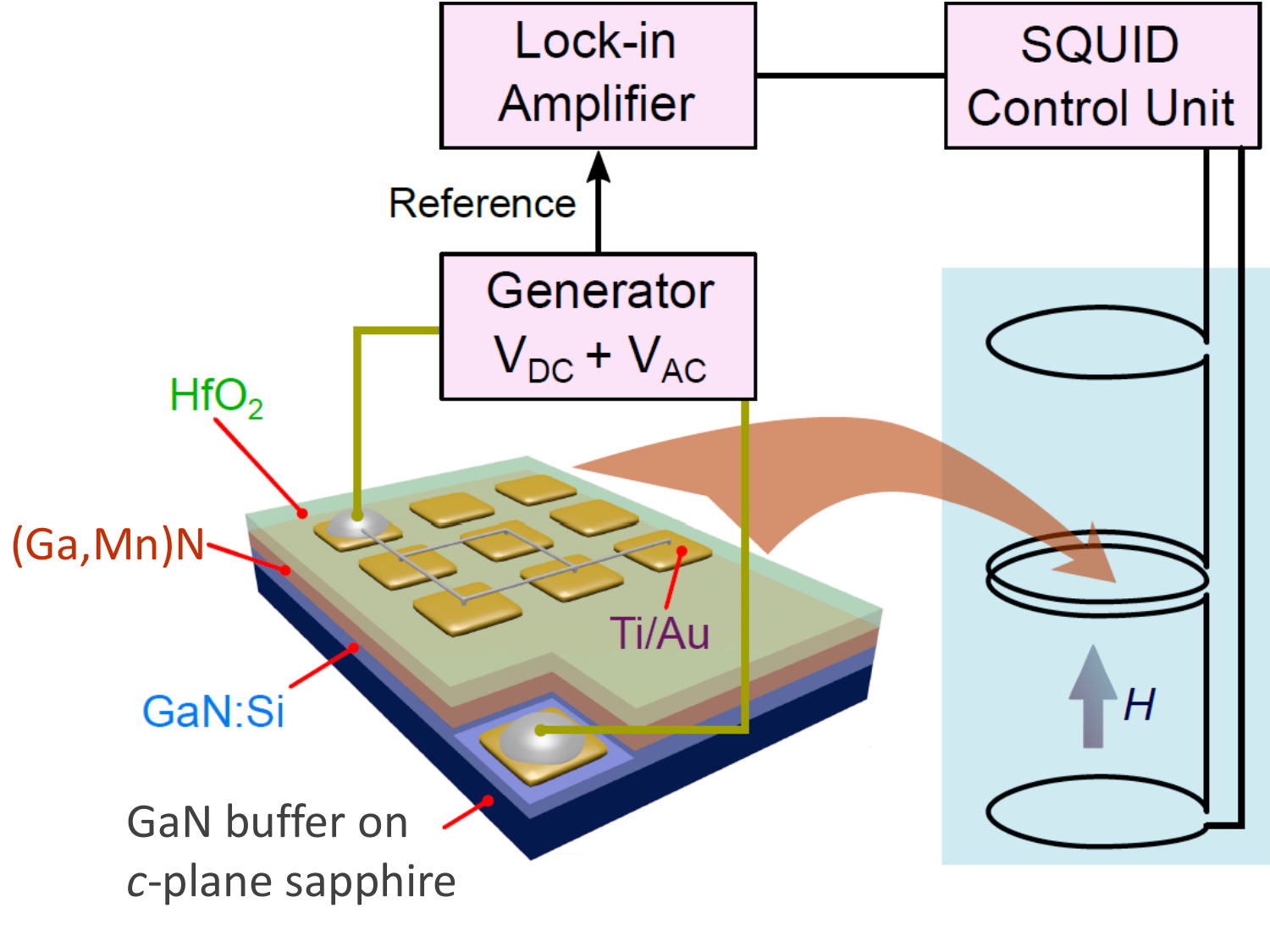}
\caption{\label{Fig:Meas_Setup} {\bf Schematic diagram of the experimental setup to perform magnetoelectric measurements in SQUID magnetometer.} The sample is located at the center of the magnetometer's pick up coils and the electrical gate induced changes of magnetic moment of (Ga,Mn)N are detected by phase sensitive fashion using a Lock-in amplifier.}
\end{figure}

For the gate-dependent measurements Ga$_{1-x}$Mn$_x$N layers need to be embedded in metal-insulator-semiconductor (MIS) structures.
As the back gate electrode we use the conductive GaN:Si top buffer layer.
To get an access to it we open by reactive ion etching (RIE) at least two $1 \times 1$\,mm$^2$ windows in the corners  of our $4 \times 5$\,mm$^2$ specimens (marked "W" in Fig.~\ref{Fig:M_sample}).
After placing photoresistive masking over these windows the whole surface is covered with 60~nm gate oxide (HfO$_2$) grown by atomic layer deposition.
Next, after re-opening of the windows by a lift-off process, we lithographically pattern between 20 to 60 of separated gold pads over the surface covered by the gate dielectric.
At the same process we deposit Au in the window areas to make the electrical contacts the GaN:Si layer -- the back gate electrode(s).
The gold pads are of an area about $0.2 \times 0.2$~mm$^2$ each and they form top capacitor plates to the back gate GaN:Si conductive layer.
Figure~\ref{Fig:M_sample} shows one of the investigated samples right after completing the patterning process.
The capacitors are then electrically tested individually at room temperature.
Only capacitors characterized by a high breakdown voltage $V_G \geq 20$~V are connected in parallel using  silver-filled electrically conductive epoxy.
A resulting total capacitance of the investigated structures varies between 0.5 to 1.2~nF.
A simplified general layout of the investigated structures is given in the central part of Fig.~\ref{Fig:Meas_Setup}.

The completed capacitor-like structure is mounted on a modified sample holder provided by Quantum Design for performing electrical measurements in their MPMS magnetometers.
In particular the modifications allows for both in plane and perpendicular orientation of the sample for electrical measurements.
We perform the magnetoelectric measurements following the protocol described in ref.~\onlinecite{Sztenkiel:2016_NatComm} with a number of small practical improvements.

The schematic diagram of the experimental set up is given in Fig.~\ref{Fig:Meas_Setup}.
Contrary to the standard method of SQUID magnetometry in which the sample is transported up and down across the SQUID pick-up coils, in our approach the sample is kept stationary at the central location of the SQUID's 2nd order gradiometer.
Having the sample stationary in SQUID a time dependent response is provided by the application of an AC modulation voltage to the gate (capacitor) over the (Ga,Mn)N layer.
The AC voltage $V_{AC}$ of an acoustic frequency $f$ ($3 \leq f \leq 17 $~Hz) via the inverse piezoelectric effect modulates the magnetic properties of the (Ga,Mn)N layer, and these changes are picked up by SQUID coils.
After amplification by the SQUID electronics, these signals  are fed to the lock-in amplifier, where the amplitude of the changes is established in the phase sensitive fashion.
We apply both the $DC$ and $AC$ external voltages: $V_{DC}$ to change the magnetic anisotropy and, usually much smaller, $V_{AC}$ to detect the resulting $M_E$.
The method allows for detection changes down to $10^{-10}$~emu, however on an expense of progressively longer acquisition times.
At $f=17$~Hz of $AC$ modulation it takes about 30~mins to complete each experimental point, including 5 minutes of initial delay to stabilize the field flux in the bore of the magnetometer's superconducting magnet.
Using  about 0.1 sec delay time during the data acquisition we collect about 12000  single readouts.
By plotting a histogram of the sample data, we obtain the magnetoelectric response, $M_E$, as the mean value of a fitted normal distribution.
It should be emphasized that this technique does not sense time independent magnetic properties of the substrate or other elements of the whole set-up.
The in-phase $AC$ magnetic response is generated only in (Ga,Mn)N located between the plates of the selected individual capacitors.

\subsection*{Density functional theory calculations}
The first-principles spin-polarized calculations are performed using a plane-wave basis set and the  Perdew-Burke-Ernzerhof exchange-correlation functional, as implemented in the Quantum-ESPRESSO package\cite{Giannozzi:2009_JPCM}. The non-collinear magnetism is taken into account by performing calculations that include spin-orbit coupling and fully relativistic norm-conserving pseudopotentials. The plane-wave cutoff is set to be 50~Ry. Five different unit cells are used, namely $2 \times 2 \times 1$, $2 \times 2 \times 2$, $2 \times 2 \times 3$, $3 \times 3 \times 2$, and $3 \times 2\sqrt{3} \times 2$ with 16, 32, 48, 72, and 96 atoms, respectively. In each case one Ga ion is replaced with Mn and the positions of all ions and are fully relaxed. For each unit cell, the lattice parameter $a$ is fixed to the theoretical value for GaN (i.e., to the substrate), and the $c$ lattice parameter is fully optimized. The structural optimization is done using ultrasoft pseudopotentials and a 30~Ry kinetic energy cut-off (240~Ry for the electronic density). For the Brillouin zone sampling $8 \times 8 \times 8$ and $3 \times 3 \times 3$  \textit{k}-point meshes are used for the smallest and largest unit cells, respectively. The calculated equilibrium values of lattice parameters for GaN are $a=0.3217$~nm and $c=0.5241$~nm, and agrees well with earlier theoretical reports\cite{Gonzalez:2011_PRB}.

\subsection*{Atomistic spin model simulations}
The magnetic energy of localized spin vectors $\textbf{S}_i$ is described by the Hamiltonian,
\begin{equation}
\label{eq:Hspin}
\mathcal{H}=\mathcal{H}_{\mathrm{Z}}+\mathcal{H}_{\mathrm{Exch}}+\mathcal{H}_{\mathrm{Aniso}},
\end{equation}
with the three terms relating to the Zeeman energy
$\mathcal{H}_{\mathrm{Z}} = -\mu_S{\sum}_{i}\textbf{S}_i\textbf{H}$, the exchange
interaction energy in the Heisenberg form $\mathcal{H}_{\mathrm{Exch}}=-{\sum_{i\neq
j}}J_{i,j}\textbf{S}_i\textbf{S}_j$, and the magnetocrystalline anisotropy energy
(MAE) \cite{Edathumkandy:2022_JMMM}.  Here, $\textbf{S}_i$ is a unit vector denoting
the local magnetic moment direction at the site $i$, $\mu_S=g\mu_BS$ is an actual value
of magnetic atomic moment with a spin $S=2$, the Land\'e $g$-factor $g=2.0$, and
$\textbf{H}$ is the external magnetic field. Following
ref.~\cite{Simserides:2014_EPJ}, we parameterize the exchange energies $J_{ij}$
\textit{vs.} Mn-Mn distances $R_{ij}$ by the exponential function
$J_{ij}=J_0 e^{-R_{ij}/b}$, with $b=1.11$~$\AA$ and $J_0 \cong 52.9$~meV.

The anisotropy term
$\mathcal{H}_{\mathrm{Aniso}}$ contains both the
trigonal (uniaxial along $[001] ||\textbf{c}$) $\mathcal{H}_{\mathrm{Krig}}$ and
triaxial $\mathcal{H}_{\mathrm{JT}}$ components (refs~\onlinecite{Gosk:2005_PRB, Wolos:2004_PRB_b,
Sztenkiel:2016_NatComm, Sztenkiel:2022_JMMM, Edathumkandy:2022_JMMM}).
The uniaxial term includes the single ion magnetic anisotropy of the Mn$^{3+}$ ions,
driven by the trigonal distortion specific to the deviation from the perfect
wurtzite atoms arrangement in the two-atom basis of GaN lattice.
The triaxial term describes the influence of the Jahn-Teller deformation, which in
wurtzite crystals grown along $c$-axis is directed along three inclined out-of-plane
directions. These three Jahn-Teller centers are distorted along
$e_{\mathrm{JT}}^A=[\sqrt{2/3},0,\sqrt{1/3}]$,
$e_{\mathrm{JT}}^B=[-\sqrt{1/6},-\sqrt{1/2},\sqrt{1/3}]$ and
$e_{\mathrm{JT}}^C=[-\sqrt{1/6},\sqrt{1/2},\sqrt{1/3}]$.
It is important to remark
here that it is precisely the presence of the three energy barriers that impedes in
plane reversal of $M$ and allows formation of hysteresis loops. The anisotropy
terms take the form,
\begin{equation}
\label{eq:HTrig}
\mathcal{H}_{\mathrm{Trig}}=-\frac{1}{2}K_{\mathrm{Trig}}{\sum_{i}}\frac{1}{2}[{S}_{iz}^2-(
{S}_{ix}^2+{S}_{iy}^2 )]
\end{equation}
\begin{align}
\label{eq:HJT}
\mathcal{H}_{\mathrm{JT}}=-\frac{1}{2}K_{\mathrm{JT}}\sum_{i}{\sum_{j=A,B,C}}(\textbf{S}_i
\cdot \textbf{e}_{\mathrm{JT}}^j)^4,
\end{align}
where $K_{\mathrm{Trig}}$ and $K_{\mathrm{JT}}$ are anisotropy parameters for
uniaxial and Jahn-Teller deformations, respectively.
The trigonal anisotropy energy can be expressed in more than one way due to the
normalization condition $S_x^2 + S_y^2 + S_z^2 = 1$. The lowest-order term is
$-\frac{1}{2}K_{\mathrm{Trig}}{S}_z^2 =
-\frac{1}{2}K_{\mathrm{Trig}}(1-{S}_x^2-{S}_y^2)$. For  $ K_{\mathrm{Trig}} > 0$ the
$\textbf{z} || \textbf{c}$ axis is magnetic easy axis and the $x,y$ plane is the
hard one. Since constant term can be omitted, we have opted for the form presented
in Eq.~\ref{eq:HTrig}.

For the modelling of the magnetoelectric effect, the trigonal term is of a particular importance. It is
exactly this term of $\mathcal{H}$ which is affected by the application of the electric
field $E$ and we compute the magnitude of gate driven effects through the
relation $K_{\mathrm{Trig}}(E) \propto \xi(E)$.

Simulations of the dynamical properties of the system are based on the stochastic
Landau-Lifshitz-Gilbert (LLG) equation applied at the atomistic
level\cite{Evans:2014_JPhysCM,Evans:2015_PRB, Edathumkandy:2022_JMMM}, which is
given by
\begin{equation}
\label{eq:LLG}
\frac{\partial{\textbf{S}_i}}{\partial t}=- \frac{\gamma}{1+\alpha_G^2}
[\textbf{S}_i\times\textbf{H}_{\text{eff}}^{i,tot}+\alpha_G\textbf{S}_i\times(\textbf{S}_i\times\textbf{H}_{\text{eff}}^{i,tot})],
\end{equation}
where $\gamma$ is the gyromagnetic ratio and $\alpha_G = 1$ is the precession
damping term. The total effective magnetic field acting on spin $i$ is composed of
two terms $\textbf{H}_{\text{eff}}^{i,tot} = \textbf{H}_{\text{eff}}^{i} + \textbf{H}_{th}^i$
(refs~\onlinecite{Skubic:2008_JPhysCM,Evans:2014_JPhysCM,Evans:2015_PRB}). The net effective
field $\textbf{H}_{\mathrm{eff}}^i=-(1/\mu_S)\partial\mathcal{H}/\partial\textbf{S}_i$ is obtained from the derivative of the total
Hamiltonian with respect to the atomic moment $\mu_S\textbf{S}_i$. The thermal
field in each spatial dimension is represented by a normal distribution
$\mathbf{\Gamma}(t) $ with a standard deviation of 1 and mean of zero, and is given
by
\begin{equation}
\textbf{H}_{th}^i= \mathbf{\Gamma} (t) \sqrt{\frac{2 \alpha_G k_B T}{\gamma
\mu_S \Delta t}},
\end{equation}
where  $\Delta t = 5\cdot10^{-6}$~ns is the
integration time step and $T$ is temperature.

\begingroup
\squeezetable
\begin{table}[h]
\centering
\caption{\textbf{Parameters of atomistic spin model and LLG simulations.} The magnetic field
is swept from 50 to -50~kOe.}
\begin{ruledtabular}
\begin{tabular}{cccc}
Magnetic field & Number of & Number of \\
 range: from, to & initialization steps & averaging steps \\
\hline
50~kOe, 3.4~kOe & $4.3\cdot10^6$ & $4.3\cdot10^6$ \\
3.4~kOe, 550~Oe & $8.6\cdot10^7$ & $8.6\cdot10^7$ \\
550~Oe, 0 & $6.7\cdot10^8$ & $6.7\cdot10^8$ \\
0, -200~Oe & $3.6\cdot10^{9}$ & $3.6\cdot10^{9}$ \\
-200~Oe, -550~Oe & $2.0\cdot10^{9}$ & $2.0\cdot10^9$ \\
-550~Oe, -3.4~kOe & $2.6\cdot10^{8}$ & $2.6\cdot10^8$ \\
-3.4~kOe, -50~kOe & $8.6\cdot10^{6}$ & $8.6\cdot10^6$ \\
\end{tabular}
\end{ruledtabular}
\label{tab:ParametersLLG}
\end{table}
\endgroup

In atomistic spin model simulations, the parallelization of the code is implemented. We
divide the simulated region into sections, where each processor simulates specified
part of the complete system. The computations are accelerated using graphics
processing units (GPUs). We use Euler's method as the integration scheme. In order to speed up
the calculations, strict synchronization of the processors (cores) is applied every fifth step of the LLG integration. The
scaling and parallelization of the code is found to be very good. We obtain practically the same
results independent of number of nodes or cores used in simulations as well as
independent of used computer architecture (based on GPU or CPU).
As the velocity of magnetization precession depends on the
strength of the magnetic field, we assume that the initialization and
averaging steps of the LLG simulations depend on the field value,
as given in Table~\ref{tab:ParametersLLG}.


\section*{Code availability}
A computer code developed within this work for atomistic Landau-Lifshitz-Gilbert computations is available at https://codeocean.com/capsule/95309f25-52bd-4e2c-9099-fd15b562a1c8/.

\section*{Data availability}
The data that support the findings of this study are available at https://codeocean.com/capsule/95309f25-52bd-4e2c-9099-fd15b562a1c8/ in the data folder.

\ \\

\section*{References}


%



\ \\

\section*{Acknowledgments}
The work is supported by the National Science Centre (Poland) through projects MAESTRO (2011/02/A/ST3/00125 - T.D.), FUGA (2014/12/S/ST3/00549 - K.G.),  OPUS (2018/31/B/ST3/03438 - M.S.) and PRELUDIUM Bis (2021/43/O/ST3/03280 - N.G.S.). This research was also supported by the Interdisciplinary Centre for Mathematical and Computational Modelling (ICM) University of Warsaw under grants no.\,G73-7, GB77-6, G85-956 and by the Foundation for Polish Science
project "MagTop" no.\,FENG.02.01-IP.05-0028/23 co-financed by the European Union from the
funds of Priority 2 of the European Funds for a Smart Economy Program 2021-2027 (FENG).

\section*{Author contributions}
The work was planned and proceeded by discussions among  M.S. and T.D. Samples growth were performed by K.G. and D.H. whereas their $x$-ray structural characterization was carried out by J.D. SQUID magnetization measurements were accomplished by K.G. and M.S, whereas D.S. processed the samples and made SQUID studies in the electric field. ALD and RIE were performed by M.F. and T.W. Analysis and interpretation of the data were accomplished by D.S., T.D. and M.S., whereas N.G.Sz performed {\em ab initio} computations. D.S., with inputs from C.S., performed magnetization simulations using atomistic spin approach. T.D., D.S., K.G. and M.S. wrote the manuscript with inputs from all the authors.

\section*{Competing interests}
The authors declare no competing interests.

\ \\
Correspondence and requests for materials should be addressed to D.S, M.S or T.D. \\

\end{document}